\documentclass[11pt]{article}

\usepackage[utf8]{inputenc}
\usepackage{fouriernc}
\usepackage{times}
\usepackage{lineno}
\usepackage{xr}
\usepackage[sort&compress,numbers]{natbib}
\usepackage{color}
\usepackage{graphicx}
\usepackage{amsmath}
\usepackage{amssymb}
\usepackage{float}
\usepackage{geometry}
\usepackage{indentfirst}
\usepackage{pdfpages}
\usepackage[colorlinks=true,allcolors=blue]{hyperref}
\usepackage[english]{babel}
\definecolor{qiblue}{rgb}{0, 0, 1}
\definecolor{gred}{rgb}{1, 0, 0}

\newcommand*\patchAmsMathEnvironmentForLineno[1]{%
	\expandafter\let\csname old#1\expandafter\endcsname\csname #1\endcsname
	\expandafter\let\csname oldend#1\expandafter\endcsname\csname end#1\endcsname
	\renewenvironment{#1}%
	{\linenomath\csname old#1\endcsname}%
	{\csname oldend#1\endcsname\endlinenomath}}%
\newcommand*\patchBothAmsMathEnvironmentsForLineno[1]{%
	\patchAmsMathEnvironmentForLineno{#1}%
	\patchAmsMathEnvironmentForLineno{#1*}}%
\AtBeginDocument{%
	\patchBothAmsMathEnvironmentsForLineno{equation}%
	\patchBothAmsMathEnvironmentsForLineno{align}%
	\patchBothAmsMathEnvironmentsForLineno{flalign}%
	\patchBothAmsMathEnvironmentsForLineno{alignat}%
	\patchBothAmsMathEnvironmentsForLineno{gather}%
	\patchBothAmsMathEnvironmentsForLineno{multline}%
}

\begin{document}

\title{\textbf{\Large Evolution of social behaviors in noisy environments}}

\date{ }

\author
{Guocheng Wang$^{1,2}$, Qi Su$^{3,4,5}$, Long Wang$^{1,6}$, Joshua B. Plotkin$^{2,7}$\\
\footnotesize{$^{1}$Center for Systems and Control, College of Engineering, Peking University, Beijing 100871, China}\\
\footnotesize{$^{2}$Department of Biology, University of Pennsylvania, Philadelphia, PA 19104, USA}\\
\footnotesize{$^{3}$School of Automation and Intelligent Sensing, Shanghai Jiao Tong University, Shanghai 200240, China}\\
\footnotesize{$^{4}$Key Laboratory of System Control and Information 
Processing, Ministry of Education of China, Shanghai 200240, China}\\
\footnotesize{$^{5}$Shanghai Key Laboratory of Perception and Control in Industrial Network Systems, Shanghai 200240, China}\\
\footnotesize{$^{6}$Center for Multi-Agent Research, Institute for Artificial Intelligence, Peking University, Beijing 100871, China}\\
\footnotesize{$^{7}$Center for Mathematical Biology, University of Pennsylvania, Philadelphia, PA 19014, USA}
}

\maketitle

\begin{abstract}
\noindent Evolutionary game theory offers a general framework to study how behaviors evolve by social learning in a population. This body of theory can accommodate a range of social dilemmas, or games, as well as real-world complexities such as spatial structure or behaviors conditioned on reputations. Nonetheless, this approach typically assumes a deterministic payoff structure for social interactions. Here, we extend evolutionary game theory to account for random changes in the social environment, so that mutual cooperation may bring different rewards today than it brings tomorrow, for example. Even when such environmental noise is unbiased, we find it can have a qualitative impact on the behaviors that evolve in a population.  Noisy payoffs can permit the stable co-existence of cooperators and defectors in the prisoner's dilemma, for example, as well as bistability in snowdrift games and stable limit cycles in rock-paper-scissors games -- dynamical phenomena that cannot occur in the absence of noise. We conclude by discussing the relevance of our framework to scenarios where the nature of social interactions is subject to external perturbations.

\end{abstract}

\clearpage

\section*{Introduction}




Game theory describes how rational individuals make strategic decisions \cite{vonneumann2007a}. Maynard-Smith and Price introduced a dynamical process, based on the principles of evolution in a population, to describe how strategic behaviors will change over time as individuals  reproduce differentially (or imitate each other's behavior) according to their fitness \cite{Smith1973}. The past decades have seen expansive development and applications of evolutionary game theory, from biology to social sciences, economics, and machine learning \cite{bandura1963social,weibull1997evolutionary,santos2018a,tuyls2007,Bloembergen2015659,Macy20027229,Borgers19971}. Evolutionary game theory now provides guiding principles to study empirical phenomena ranging from the prevalence of prosocial behaviors in animal or human societies, to the effects of spatial and network structure on behavior, or even the behavior of individuals during an epidemic \cite{roca2009evolutionary,chang2020game,antonioni2017coevolution}.

When an individual engages in many social interactions in a population, their total payoff depends on the frequencies of all strategies. According to the central tenet of evolutionary game theory, those strategies that yield higher total payoffs tend to proliferate. In the limit of a large population, this dynamical process  can be  described by the so-called replicator equation, an ordinary differential equation for the dynamics of strategy frequencies \cite{Taylor1978,Schuster1983}.
Analysis of the replicator equation for a game, including its equilibrium points and phase portraits, provides insight into the long-term strategic outcomes. For two-strategy games, the evolutionary outcomes can be categorized into three scenarios: strategy dominance,  when one strategy will eventually overtake the population regardless of initial state, as exemplified by the prisoner’s dilemma; strategy coexistence, with a unique stable interior equilibrium so that both strategies persist in the population, as seen in the snowdrift game; and bi-stability, with an unstable interior equilibrium and both boundaries absorbing, as in coordination games. More generally, applications of evolutionary game theory include explanations for the evolution of cooperation (prisoner’s dilemma and snowdrift game), for social coordination (coordination game) \cite{Pacheco2009,Johnstone2011},  biodiversity \cite{Kerr2002,Reichenbach2007}, and  learning (rock-paper-scissors game) \cite{Littman1994}.

Social interactions involving more strategies produce a greater diversity of dynamical outcomes. For instance, in three-strategy games there can be as many as 33 distinct phase portraits  \cite{Hofbauer2003,Zeeman1980,Bomze1983,Bomze1995}. The rock-paper-scissors (RPS) game holds particular interest, because each strategy dominates one other strategy while being dominated by another, which can lead to decaying cycles among the three strategies. 
One fundamental result is that, despite this complexity, there are no isolated periodic orbits in classical replicator dynamics -- that is, no stable limit cycles, in any three-strategy game. 

Despite theoretical advances and practical applications of evolutionary game theory, most work has operated under the assumption of a fixed social environment (but see \cite{Hilbe2018a,Su2019a,Weitz2016,Tilman2020,Wang2021,Taitelbaum2020,Stollmeier2018,zhengEnvironmental2018}). This assumption posits that the payoffs associated with the various outcomes in a social interaction remain the same throughout the evolutionary process. Although this idealized assumption simplifies mathematical analyses and allows for a systematic classification of possible outcomes in two-strategy games, this framework does not account for the stochastic nature of real-world interactions, especially in open environments \cite{luhmann2011intolerance,ketchpel1994forming}. 
In reality, most interaction outcomes are subject to stochastic perturbations, stemming from externalities such as changing resource availability, economic conditions, etc. Perturbations to payoffs are often irregular, unpredictable, and they may vary in strength.

Noisy payoffs are commonplace in empirical systems, ranging from microbes to humans. In bacterial communities, for example, stochastic variation in nutrients, toxins, or temperature \cite{cooper2010experimental,nguyen2021environmental,rashit1987environmental}
shapes the energetic rewards to strains that either share or do not share diffusible resources, such as siderophores \cite{cordero2012public,Caceres1997,descamps-julienStable2005}. 
Even in human societies, fluctuations in the stock market can bring stochastic returns, which greatly affect human behaviors and economic cycles compared with risk-free (constant) returns \cite{chauvetStock1999}.
It is important, therefore, to extend evolutionary game theory to accommodate noisy social environments and to understand their impacts on long-term behavioral patterns.

The form of noise we address in this study is substantially different from demographic noise \cite{Constable2016,wang2023a,huang2015stochastic,gillespie1974natural} or from noise in payoff observations \cite{wang2024}. Under scenarios with demographic stochasticity or observation uncertainty, individuals are subject to independent perturbations in each time step. Instead of stochasticity in population size or observations, we will focus on noise in the nature of social interactions themselves, meaning a source of global perturbation that influences the payoffs associated with each possible outcome in a social interaction -- a perturbation that applies to all individuals simultaneously.


Here we extend evolutionary game theory to account for environmental noise in social interactions. To model this, the payoff structure of the game will be subject to an independent unbiased perturbation at each time step. For example, mutual cooperation between a pair of individuals may yield a slightly different payoff today than mutual cooperation will yield tomorrow; and these perturbations are shared across the entire population. In our model, strategies with higher realized payoffs tend to proliferate in each time step, in keeping with the central tenet of evolutionary game theory. We will derive the corresponding dynamical equations for strategy frequencies in such noisy environments; these are ordinary differential equations represented as a modification of the classical replicator equation \cite{Taylor1978}.
We will analyze the long-term behavior of these dynamical systems
to systematically describe all possible dynamics for two-strategy games. We find that some dynamical outcomes in two-strategy games with noisy payoffs are qualitatively different than any possible outcome under classical replicator dynamics without noise.
Shifting our focus to three-strategy games, we show that noisy games can induce stable limit cycles --- a dynamic that cannot occur without noise -- and we will delineate the conditions that produce these stable cycles. These results underscore the substantial impact of a noisy social environment on the evolution of social behaviors: noise tends to increase the diversity of behaviors.


\section*{Model}
 \begin{figure}
    \centering
    \includegraphics[scale=0.7]{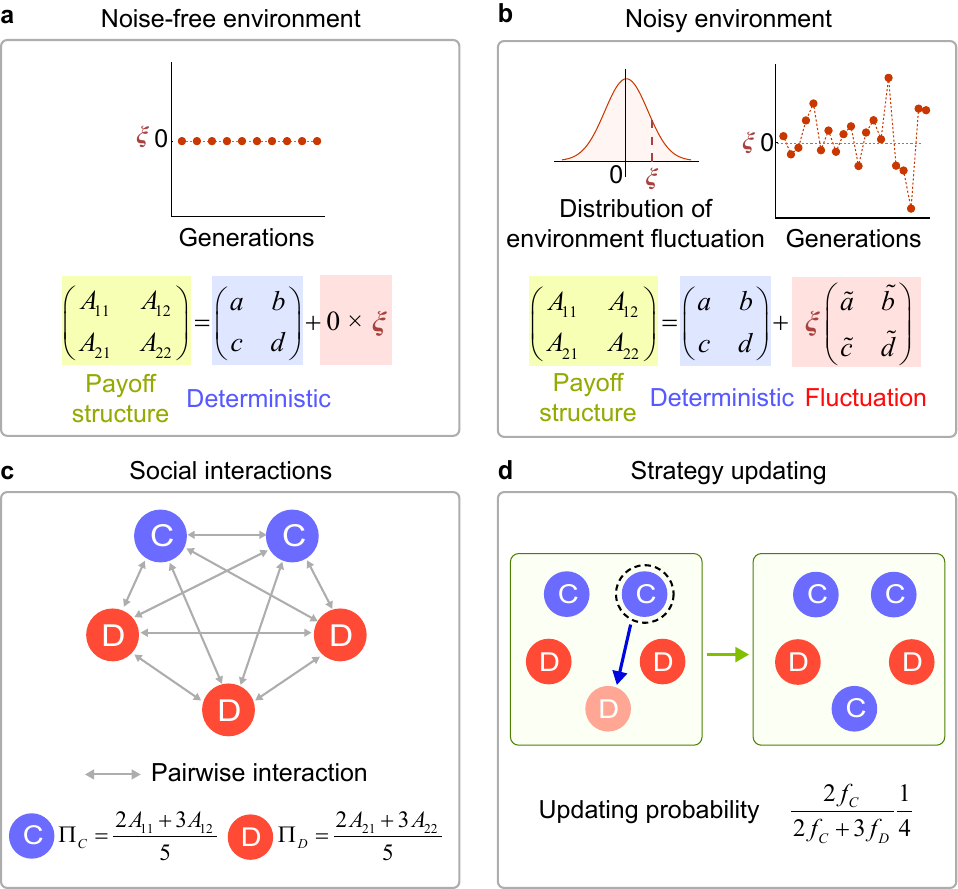}
    \caption{\textbf{Evolution of behavior in a noisy social environment.} 
    We model a noisy social environment as a payoff matrix with a deterministic component and a random component. \textbf{a}, In the noise-free setting, the payoff structure is the same in all generations. \textbf{b}, In a noisy environment, a random perturbation $\xi$ is sampled independently from the standard Gaussian distribution each generation. 
    \textbf{c}, In each generation, every individual interacts pairwise with all others and derives an average payoff, i.e.~$\Pi_C$ for cooperators and $\Pi_D$ for defectors, which depends on the current payoff matrix. 
    The figure illustrates an example of $\Pi_C$ and $\Pi_D$ in a population consisting of $2$ cooperators and $3$ defectors.
    Each individual's payoff $\Pi$ determines their fitness according to $f=\text{exp}(s\Pi)$.
    \textbf{d}, Following all social interactions, an  individual is selected to serve as the role model, chosen randomly proportional to fitness (dashed circle), and their strategy is imitated by another individual chosen uniformly (birth-death updating).
    }
    \label{fig1}
\end{figure}

We consider a well-mixed population of $N$ individuals engaged in pairwise social interactions. Focusing first on two-strategy games, each individual employs a behavioral strategy, generically denoted as either cooperate ($C$) or defect ($D$), for interactions with all other individuals in a $2 \times 2$ symmetric game. (This terminology has a natural interpretation when the payoff matrix encodes a prisoner's dilemma, but we nonetheless retain the same terminology for the two strategies in any $2\times 2$ game, including cases where $C$ is no more ``cooperative'' than $D$.) The payoffs received by a pair of interacting individuals are described by the following matrix, comprised of both a deterministic and a stochastic component:
\begin{equation}
    \begin{bmatrix}
    A_{11} & A_{12} \\A_{21} &A_{22}
    \end{bmatrix}=\begin{bmatrix}
    a & b \\c &d
    \end{bmatrix}+\xi \begin{bmatrix}
    \tilde{a} & \tilde{b} \\ \tilde{c} & \tilde{d}
    \end{bmatrix},
    \label{eq:payoff matrix}
\end{equation}
where $\xi$ is a random variable. More general forms of noise are considered in Section 3.2 of Supplementary Information.

The deterministic portion of the payoffs in Eq.~\ref{eq:payoff matrix} remains constant throughout the evolutionary process of payoff-biased imitation. Whereas the stochastic component includes a noise element $\xi$ that is sampled from the standard Gaussian distribution independently each generation. The amount of noise is modulated by a time-invariant amplitude matrix, which depends on the strategies of the two players. Importantly, the noise term $\xi$ is sampled once per generation and it applies to all social interactions in that generation, which captures the idea of global environmental perturbations in the reward structure. For example, mutual cooperation may be more beneficial at one time than it is at another time. The extent of perturbation for certain pairs of strategies (e.g., mutual cooperation, $\tilde{a}$) may differ than for other pairs of strategies (e.g., mutual defection, $\tilde{d}$).

According to the payoff matrix above, mutual cooperation brings each individual a (random) payoff $A_{11}$ ($a+\xi\widetilde{a}$) and mutual defection brings each individual $A_{22}$ ($d+\xi\widetilde{d}$); whereas 
unilateral cooperation brings the cooperator $A_{12}$ ($b+\xi\widetilde{b}$) and the defector $A_{21}$ ($c+\xi\widetilde{c}$). After all pairwise interactions in the population, each individual obtains an average payoff. 
In a population with $n_C$ cooperators and $N-n_C$ defectors, a cooperator and defector respectively receive average payoffs
\begin{subequations}
\begin{align}
    \Pi_C=\frac{1}{N}\left(n_CA_{11}+(N-n_C)A_{12}\right), \\
    \Pi_D=\frac{1}{N}\left(n_CA_{21}+(N-n_C)A_{22}\right).
\end{align}   
\end{subequations}
The payoff $\Pi$ is then transformed into an individual's fitness $f$ according to the function $f=\text{exp}(s\Pi)$.
The parameter $s$ is called the strength of selection, and it measures to what degree the outcomes of social interactions influence the proliferation of behavioral types  (either through literal births and deaths, or by payoff-biased imitation) \cite{Wu2010selection,Wu2013}.

After all pairwise games are played, an individual in the population updates their strategy according to a birth-death updating rule, which can alternatively be seen as payoff-biased strategy imitation. 
That is, a random individual is selected to reproduce (or to serve as the role model) with probability proportional to his fitness; and his offspring replaces another individual (or imitator) selected uniformly at random. 
For example, the probability that individual $i$ reproduces and replaces (equivalently, is imitated by) individual $j$ is given by
\begin{equation}
e_{ij}=\frac{f_i}{\sum_{\ell = 1}^N f_{\ell}}\frac{1}{N-1},
\end{equation}
where $f_{\ell}$ denotes the fitness of individual $\ell$. 
The exponential fitness function, which is widely used in literature \cite{Nowak2006a,Wu2010selection,Wu2013}, guarantees positive fitness 
even for strong selection. It also has a nice property that the resulting dynamics under a noisy environment depend only on differences among fluctuation intensities (i.e., differences among $\tilde{a}\sim \tilde{d}$, see below), analogous to the classic replicator equation whose dynamics depend only on payoff differences among strategies.

\section*{Results}
The only absorbing states for the evolutionary process above are populations consisting of all cooperators or all defectors. A population starting from any initial configuration will eventually converge to one of these two states, through a combination of selection induced by deterministic payoff differences, noise in payoffs, and demographic stochasticity. We have analyzed the joint effects of all three phenomena by studying the fixation probability and mean fixation time  (see Supplementary Fig.~1 and Section 3.3 in Supplementary Information). In the main text, however, we will focus our analysis on the limit of large populations. More precisely, we focus on the regime $s \ll \sqrt{1/N}$, where the effects of demographic noise can be neglected (see Section 2.1 of Supplementary Information).
This allows us to pinpoint the effects of noise in the environment alone, and it also provides information about the dynamics on the interior of state space, when multiple types are present in the population. 
We begin our investigation with two-strategy games; 
subsequently we extend our analysis to encompass multiple-strategy games, including arbitrary numbers of strategies and different forms of environmental noise.
Finally, we undertake a mathematical analysis of general fitness functions and rules for updating strategies.

\subsection*{Behavioral evolution in two-strategy games}

\begin{figure}
    \centering
    \includegraphics[width=\textwidth]{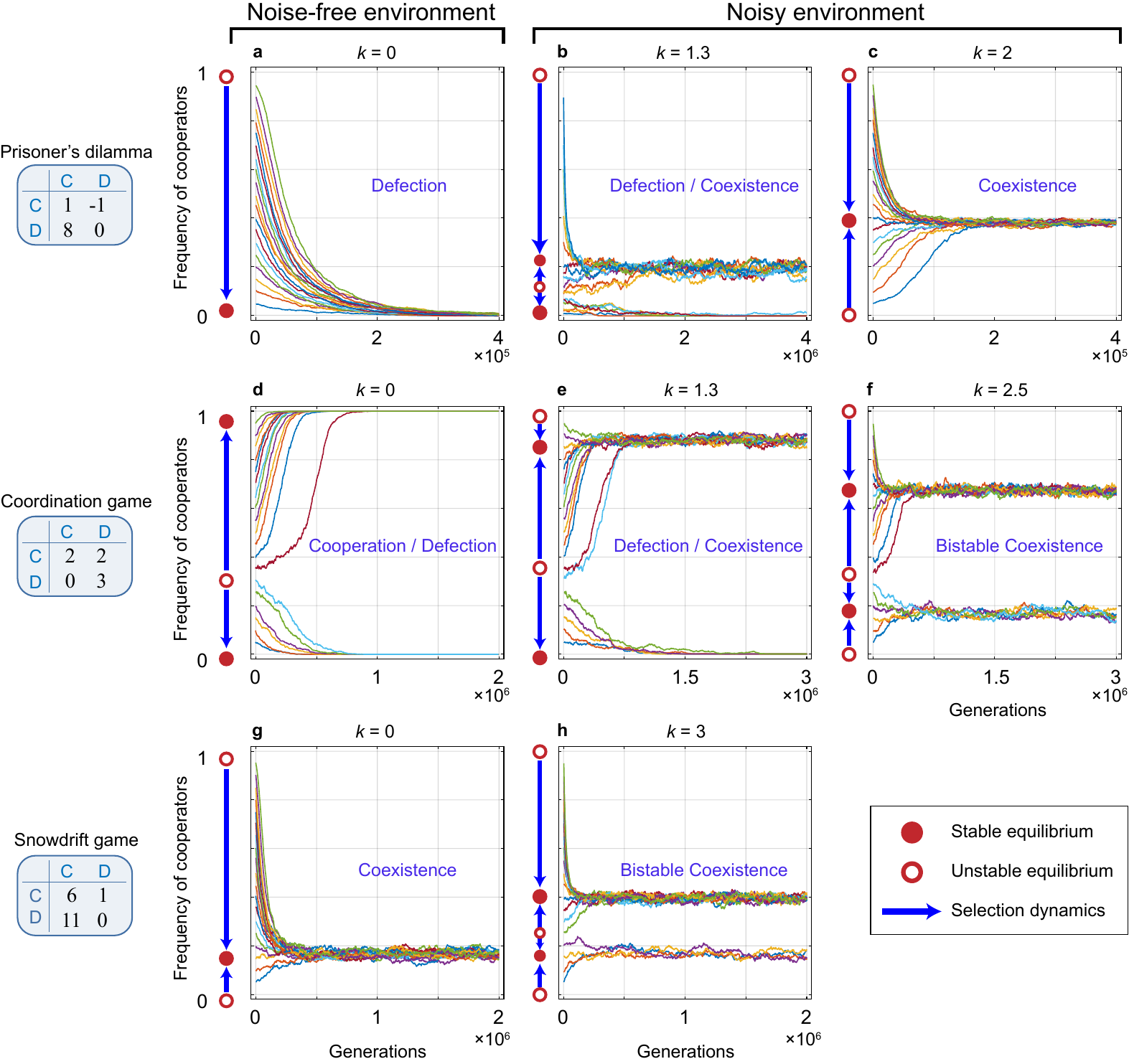}
    \caption{
    \textbf{Behavioral outcomes in noisy social environments.} The figure illustrates evolutionary dynamics for three types of games (prisoner's dilemma, coordination, and snowdrift games) in either a noise-free environment  (panels \textbf{adg}) or a noisy environment with different fluctuation intensities (panels \textbf{bcefh}). Each panel shows several evolutionary trajectories starting from various initial states; and the $y$-axis also indicates the direction of evolution as well as unstable (open circles) and stable (solid circles) equilibria. Environmental fluctuations increase the diversity of long-term outcomes. In the prisoner's dilemma, for instance, the unique dominance of defection in the noise-free environment (\textbf{a}) can be augmented with a new equilibrium containing a mixture of cooperation and defection (\textbf{b}), or, if the fluctuation intensity is larger, coexistence may become the only possible outcome (\textbf{c}). For other games, noisy environments can produce new qualitative outcomes such as defection/coexistence (\textbf{e}) or bistable coexistence (\textbf{f,h}) that cannot occur in the absence of noise. Parameters: $s=0.1$, $N=10000$.
    }
    \label{fig2}
\end{figure}

For an arbitrary symmetric $2\times 2$ game, we generically denote the two possible strategies as cooperate ($C$) and defect ($D$). 
Let $x=n_C/N$ denote the frequency of cooperators in the population and $1-x$ the frequency of defectors. The deterministic portion of the average payoffs for cooperators and defectors are then $\pi_C=ax+b(1-x)$ and $\pi_D=cx+d(1-x)$ respectively. Under the standard assumption that the selection intensity is weak ($s\ll 1$) \cite{Ohtsuki2006,Ohtsuki2006replicator,Wu2010selection,Allen2017}, the evolution of strategy frequencies in the presence of environmental noise can be described by the following ordinary differential equation (for a derivation see Methods and Section 2.1 in Supplementary Information)
\begin{equation}
    \dot{x}=sx(1-x)\left[\pi_C-\pi_D+s\left(\frac{1}{2}-x\right)\left(\tilde{b}-\tilde{d}+x(\tilde{a}-\tilde{b}-\tilde{c}+\tilde{d})\right)^2\right].
    \label{eq:system equation original}
\end{equation}
This equation reduces to the classical replicator equation in a setting without any environmental noise (i.e.~$\tilde{a}=\tilde{b}=\tilde{c}=\tilde{d}=0$). Moreover, if all elements of the payoff matrix experience noise of the same magnitude (i.e., $\tilde{a}=\tilde{b}=\tilde{c}=\tilde{d}$), then again this equation simplifies to the classic replicator equation --  which means that environmental fluctuations that perturb all payoffs equally have no effect on the evolutionary dynamics of strategic types.

To analyze the dynamics in the presence of nontrivial noise structures, we will assume that the deterministic portion of the payoffs in Eq.~\ref{eq:payoff matrix}, namely $a,b,c,d$, are all of order $O(1)$. If all elements in the stochastic component of payoff structure (namely, ~$\tilde{a},\tilde{b},\tilde{c},\tilde{d}$) are much smaller than $1/\sqrt{s}$, then the term $\left(\tilde{b}-\tilde{d}+x(\tilde{a}-\tilde{b}-\tilde{c}+\tilde{d})\right)^2$ in Eq.~\ref{eq:system equation original} is negligible compared to the term $\pi_C-\pi_D$, so that the deterministic payoff dominates and the dynamics will correspond to the classic replicator equation. 
On the other hand, if there exists at least one element in the fluctuation part that is much larger than $1/\sqrt{s}$, then the term $\pi_C-\pi_D$ is negligible compared to the term $\left(\tilde{b}-\tilde{d}+x(\tilde{a}-\tilde{b}-\tilde{c}+\tilde{d})\right)^2$, so that noise alone governs the direction of strategic evolution.
Therefore, we will hereafter focus on the regime in which fluctuations are all of order of $1/\sqrt{s}$, so that terms arising from both deterministic and stochastic components have non-negligible effects on the evolutionary dynamics.
Rescaling the noisy contributions by 
\begin{equation}
    \begin{bmatrix}
    \sigma_a & \sigma_b \\ \sigma_c & \sigma_d
    \end{bmatrix}=\sqrt{s}\begin{bmatrix}
    \tilde{a}&\tilde{b} \\ \tilde{c}&\tilde{d}
    \end{bmatrix},
    \label{eq:rescale_noise}
\end{equation}
and introducing the notation $\sigma_C=\sigma_{a}x+\sigma_{b}(1-x)$ and $\sigma_D=\sigma_{c}x+\sigma_{d}(1-x)$, we can rewrite Eq.~\ref{eq:general system equation} as
\begin{equation}
        \dot{x}=sx(1-x)\left[\pi_C-\pi_D+\left(\frac{1}{2}-x\right)(\sigma_C-\sigma_D)^2\right].
        \label{eq:replicator}
\end{equation}
We note that replicator equation is derived under the assumption of weak selection. In Supplementary Information we show that the replicator equation nonetheless provides a good approximation even for moderate and strong selection  (see Supplementary Fig.~2). Moreover, for strong selection ($s \sim O(1)$), noise needs only to have the same (or larger) magnitude as mean payoffs (i.e. $1/\sqrt{s}\sim O(1)$) to influence evolutionary dynamics, which is a relaxed condition compared to the case of weak selection. 

By varying $\sigma_a$, $\sigma_b$, $\sigma_c$, and $\sigma_d$ in Eq.~\ref{eq:rescale_noise}, we can model various form of environmental fluctuations. We can use Eq.~\ref{eq:replicator} to explore the resulting evolutionary dynamics in a given noisy environment. Here we focus on a representative scenario where the noise intensity is directly proportional to the deterministic portion of the payoff, as described by the equation
\begin{equation}
    \begin{bmatrix}
    \sigma_a & \sigma_b \\ \sigma_c & \sigma_d
    \end{bmatrix}=k\begin{bmatrix}
    a&b \\c&d
    \end{bmatrix}.
    \label{eq:proportional noise}
\end{equation}
This assumption captures the idea that the relative fluctuations of all payoffs are the same, although the absolute fluctuations may differ (and, indeed, must differ for any departure from classical behavior). 
In the absence of noise, Eq.~\ref{eq:replicator} reduces to the classic replicator equation, which exhibits at most three equilibria: $x=0$, $x=1$, and $x=x^*=(d-b)/(a-b-c+d)$ (the special case of $a-b-c+d=0$ is discussed in Section 2.1 of Supplementary Information). 
As a result, all two-strategy games without noise can be categorized into four dynamical outcomes: Prisoner's dilemma type, where the equilibria are $x=0$ and $x=1$ 
and only $x=0$ (full defection) is stable (Note that we generically call any defection-dominant game a prisoner's dilemma, even though it may not be a social dilemma, (i.e., $2a>b+c$ and $a>d$ are not required));
Harmony type games, where the equilibria are $x=0$ and $x=1$, and only $x=1$ (full cooperation) is stable;
Coexistence games (e.g., snowdrift game), where all three equilibria are within $[0,1]$, and only $x=x^*$ (coexistence of types) is stable;
Coordination games, where all three equilibria are within $[0,1]$, and only $x=0$ and $x=1$  are stable.
These dynamics and simulated Monte Carlo trajectories for $N$ finite are illustrated in Fig.~\ref{fig2}a, d, g (the harmony game is omitted as it is equivalent to a prisoner's dilemma by swapping the names of the strategies).

In contrast to the classical setting summarized above, environmental fluctuations (Eq.~\ref{eq:replicator}) induce up to five different equilibria in the interval $[0,1]$, which enriches the diversity of long-term behavioral outcomes.  In the prisoner's dilemma, for example, defectors will eventually dominate the population in a noise-free environment (Fig.~\ref{fig2}a), but environmental fluctuations can promote cooperation and lead to a stable coexistence of defectors and cooperators (Fig.~\ref{fig2}b,c), which is reminiscent of the outcome classically seen in a (noiseless) coexistence game.
Furthermore, in coexistence and coordination games, noisy environments can produce bistable dynamics with one equilibrium on the boundary and one in the interior (Fig.~\ref{fig2}e); and they can even produce two stable interior equilibria (Fig.~\ref{fig2}f, h), meaning different stable mixtures of the two strategic types can both be stable. These dynamical outcomes cannot occur in classical settings without noise. As these results show, noisy social environments enrich the space of behavioral outcomes, by destabilizing boundary equilibria while increasing the number of stable interior equilibria that support the coexistence of types.

\begin{figure}
    \centering
    \includegraphics[width=\textwidth]{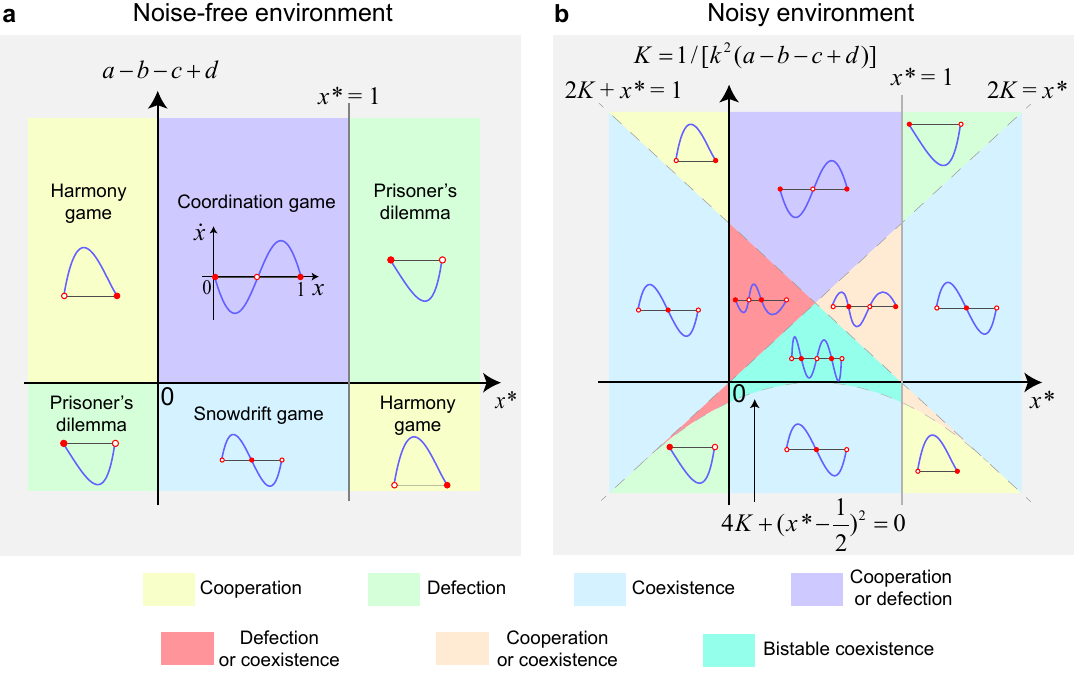}
    \caption{\textbf{Classification of all behavioral outcomes for $2\times2$ games in noisy and noise-free social environments.}
    Two key variables, $x^*=(d-b)/(a-b-c+d)$ and $K=1/[k^2(a-b-c+d)]$, which depend on the deterministic payoff matrix and the intensity of noise, effectively categorize all possible dynamical outcomes.  Insets display how the change in cooperator frequency, $\dot{x}$, depends on the current frequency of cooperators, $x$, with solid (open) circles denoting stable (unstable) equilibria.
    {\bf{a}}, In the absence of noise ($k=0$) there are four distinct dynamical outcomes that depend only on the sign of $K$ and the value of $x^*$. For example, games in the regime $x^*<0$ and $K<0$ (bottom-left region, green) have a unique outcome dominated by defection. 
    {\bf{b}}, In the presence of noise ($k>0$) there are seven possible dynamical outcomes, three of which cannot occur without environmental noise (i.e., $k=0$): an interior stable equilibrium and a stable equilibrium on the boundary $x=0$ (red region), an interior stable equilibrium and a stable equilibrium on the boundary $x=1$ (peach region), and two interior stable equilibria (teal region). We have derived an analytic classification of which dynamical pattern will arise, determined by lines and parabolas in the parameters $K$ and $x^*$, as indicated in the figure.
    } 
    \label{fig3}
\end{figure}

We can systematically classify all dynamical patterns for $2 \times 2$ games.
For the space of environmental perturbations described by Eqs.~\ref{eq:replicator} and \ref{eq:proportional noise}, the long-term behavioral outcomes are determined by two compound parameters: $x^*=(d-b)/(a-b-c+d)$ and $K=1/(k^2(a-b-c+d))$ (see Section 2.1.2 in Supplementary Information).
The parameter $x^*$ represents an equilibrium point of the classic replicator equation, and the parameter $K$ involves both the deterministic component of payoffs and the overall amount of noise ($k$). 
Whereas there are four possible dynamical outcomes for a noise-free environment (Fig.~\ref{fig3}a), we find that seven distinct dynamical outcomes can arise in noisy environments (Fig.~\ref{fig3}b). 
In particular, there are at most five equilibrium points in $[0,1]$, and their number and stabilities are determined by the values of $x^*$ and $K$. 
In fact, we have an analytical understanding which dynamical outcome will occur, which is determined by the values of $x^*$ and $K$ (Fig.~\ref{fig3}b).

To illustrate how to use our classification of long-term outcomes we consider an example game with deterministic payoff matrix $[1,-1;8,0]$ (the same game that was depicted in Fig.~\ref{fig2}a). In the classical setting without noise ($k \rightarrow 0$), the internal equilibrium point is $x^*=-1/6$, and the sign of $K$ is negative, which corresponds to a prisoner's dilemma (the bottom-left region in Fig.~\ref{fig3}a) with a unique stable equilibrium at $x=0$, meaning pure defection. 
However, as the intensity of noise ($k$) increases, the point $(x^*, K)$ passes through three different regions in Fig.~\ref{fig3}b, leading to three types of long-term outcomes: defection, defection or co-existence, and assured coexistence. Indeed, we previously observed these three distinct outcomes, empirically, in Fig.~\ref{fig2} (panels abc) as we changed the level of noise. As this example illustrates, the classification scheme summarized in Fig.~\ref{fig3} provides a synthetic and analytical way to predict how the deterministic and noisy components of payoffs collectively determine long-term behavioral outcomes in a population.

It is worth noting that we have focused on the case when the noise intensity is proportional to the deterministic payoff (Eq.~\ref{eq:proportional noise}), which produces one of seven distinct dynamical outcomes determined by the value of $(x^*,K)$. But in fact there remain only seven distinct dynamical outcomes even for arbitrary variance structures ($\sigma_a, \sigma_b, \sigma_c, \sigma_d$), as shown in Fig.~\ref{fig3}b (see Section 2.1.2 Supplementary Information).

\subsection*{Behavioral evolution in multi-strategy games}

We can extend evolutionary game theory to account for noisy environments when social interactions involve more than two strategies, denoted by a strategy set $\mathcal{L}=\{1,2,\cdots, \ell\}$.
The general payoff structure of such a game is given by  
\begin{equation}
    A_{ij}=a_{ij}+\frac{\xi}{\sqrt{s}}\sigma_{ij},
    \label{eq: general payoff elements}
\end{equation}
where $A_{ij}$ represents the payoff derived by an individual taking strategy $i$ against the opponent with strategy $j$. As before, the payoff is again comprised of a deterministic component $a_{ij}$ and a stochastic component $\xi\sigma_{ij}/\sqrt{s}$. 
If we let $x_i$ denote the frequency of individuals in a population using strategy $i$, and $\pi_i=\sum_{j}a_{ij}x_j$ denote the deterministic component of payoff to strategy  $i$, and $\sigma_i=\sum_{j}\sigma_{ij}x_j$ denote the fluctuation component of the payoff, then the evolutionary dynamics for multi-strategy games in noisy environments are given by
\begin{equation}
    \dot{x}_i=sx_i\left[\pi_i-\bar{\pi}+\frac{1}{2}\left((\sigma_i-\bar{\sigma})^2-\sum_{i}x_i(\sigma_i-\bar{\sigma})^2 \right) \right].
    \label{eq:multi-strategy system equation}
\end{equation}
Here $\bar{\pi}=\sum_{i}\pi_ix_i$ denotes the average deterministic component of payoff over the population and $\bar{\sigma}=\sum_{i}\sigma_ix_i$ denotes the average fluctuation intensity.
Eq.~\ref{eq:multi-strategy system equation} reduces to the classic replicator equation in the absence of environmental noise ($\sigma_{ij}=0$).


\begin{figure}
    \centering
    \includegraphics[width=\textwidth]{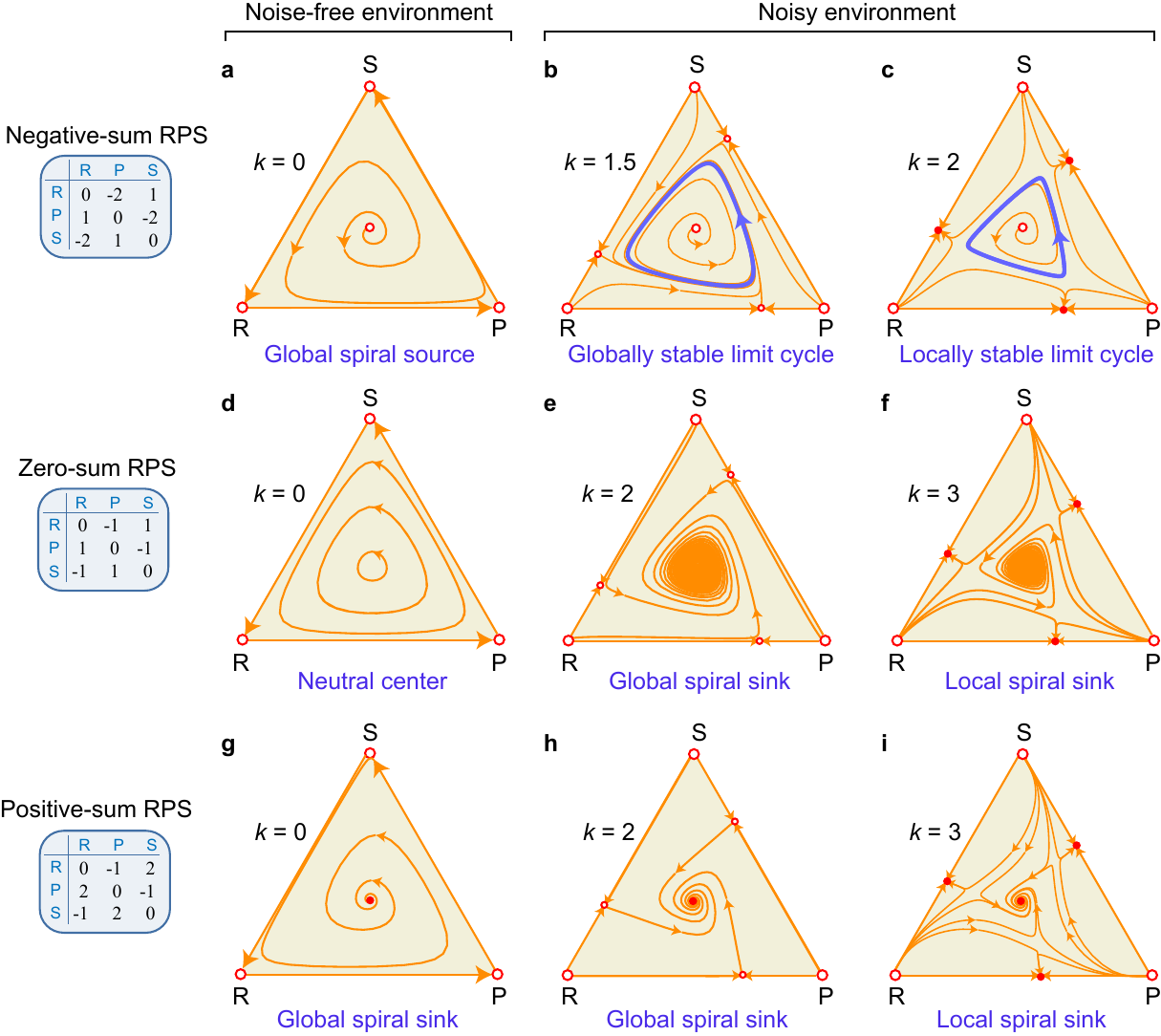}
    \caption{
    \textbf{Evolution of behavior in Rock-Paper-Scissors games.}
In the classical setting without noise, the dynamical pattern of behaviors in a rock-paper-scissors (RPS) game hinges on the relative magnitudes of payoffs $\alpha$ (paper vs rock) and $\beta$ (rock vs scissors). {\bf{a}}, For $\alpha>\beta$, the interior equilibrium with a mixture of types ($x=y=z=1/3$) is an unstable focus, with all trajectories spiralling out towards the boundary.
{\bf{d}}, In the case  $\beta=\alpha$, the interior equilibrium becomes a center, and all trajectories manifest as isolated closed orbits, which are unstable.
{\bf{g}}, For $\alpha<\beta$, the interior equilibrium becomes a stable focus inducing a spiral sink towards the mixture of all three types.
Introducing environmental noise produces a variety of new dynamical phenomena. For $\alpha>\beta$, noise of intermediate intensity can induce a stable limit cycle ({\bf{b}}) which is globally attractive: all trajectories ultimately converge to this orbit. When the intensity of noise is yet stronger, there is another stable limit cycle ({\bf{c}}), although it is not globally attractive and it co-occurs with three stable equilibria each featuring a mixture of two strategies.
In scenarios when $\alpha=\beta$ or $\alpha<\beta$ environmental noise can render the interior equilibrium asymptotically stable ({\bf{ef}}) and increase its stability ({\bf{hi}}). As the intensity of noise increases, trajectories converge more rapidly towards the interior equilibrium, accompanied by the emergence of new equilibria with stable mixtures of two strategies.
Parameters: $s=0.1$.
    }  
    
    \label{fig5}
\end{figure}

We will use Eq.~\ref{eq:multi-strategy system equation} 
to study behavioral evolution with noise for the most famous three-strategy game, namely the rock-paper-scissors (RPS) game \cite{Hofbauer2003,Zeeman1980,Bomze1983,Bomze1995}. The payoff structure can be represented as:
\begin{equation}
    \begin{bmatrix}
    0&-\alpha&\beta \\
    \beta &0 &-\alpha \\
    -\alpha & \beta & 0
    \end{bmatrix},
\end{equation}
where $\alpha>0$ and $\beta>0$, and the three strategies (``rock'', ``paper'', and ``scissors'') correspond to each row and column in order.
In a classical noise-free setting, the three strategies — ``rock'', ``paper'', and ``scissors'' — dominate each other cyclically, producing either decaying cycles towards a stable mixture of all three strategies, oscillation among states dominated by one strategy then another, or neutral oscillations containing a mixture of strategies (Fig.~\ref{fig5}a,d,g). 
In particular, the point $e^*=(1/3,1/3,1/3)$ is the only interior equilibrium, and its stability depends on the relative magnitude of $\alpha$ and $\beta$. For $\alpha>\beta$ (called negative-sum RPS games), $e^*$ is a spiral source (see Fig.~\ref{fig5}a) and all trajectories converge to the boundary spirally. For $\alpha=\beta$ (called zero-sum RPS), $e^*$ is a neutral center and all trajectories are closed orbits (see Fig.~\ref{fig5}d). And for $\alpha<\beta$ (called positive-sum RPS), $e^*$ is a spiral sink and all trajectories converge to $e^*$ spirally (see Fig.~\ref{fig5}g). 


The presence of environmental noise qualitatively changes these dynamical outcomes.
As before, we consider environmental fluctuations whose intensity is proportional to the deterministic payoff, i.e., $\sigma_{ij}=ka_{ij}$. We find that when $\alpha$ is close to $\beta$, there exist stable limit cycles, which cannot possibly occur in a noise-free environment, except in the presence of mutations \cite{mobiliaOscillatory2010}.
In particular, for negative-sum PRS games (i.e., $\alpha>\beta$) environmental fluctuations can induce a stable limit cycle (see Fig.~\ref{fig5}b, c) and this limit cycle is globally stable provided the fluctuation intensity is not too strong (Section 2.2 in Supplementary Information):
\begin{equation}
    |k|<\frac{\sqrt{2}(\alpha+2\beta)}{\sqrt{\alpha (\alpha^2+\beta^2+\alpha\beta)}}.
    \label{eq:global_stable}
\end{equation}
In this regime, trajectories starting from any initial composition of strategies will converge to a closed, stable orbit (Fig.~\ref{fig5}b).
The intensity of fluctuations influences the amplitude of this limit cycle. As $k$ increases, the diameter of the limit cycle becomes smaller, which implies a smaller amplitude of oscillation around the interior point $e^*$. For $k$ yet larger than the condition given by Eq.~\ref{eq:global_stable} the limit cycle is still stable but it is not globally attractive because there are also some stable equilibria points on the boundary containing mixtures of two strategies (Fig.~\ref{fig5}c).
In the case of zero-sum PRS games, i.e., $\alpha=\beta$, trajectories may either be a very slow global sink (Fig.~\ref{fig5}e) or, if noise is large enough, they will alternatively converge to either the interior point $e^*$ or to one of three stable equilibria on the boundary that consists of a mixture of two strategies (Fig.~\ref{fig5}f). For positive-sum PRS games, i.e., $\alpha<\beta$, the condition that $e^*$ is globally stable is the same as Eq.~\ref{eq:global_stable}, but trajectories converge to $e^*$ more rapidly than in the absence of noise (Fig.~\ref{fig5}h). Alternatively, if the intensity of noise is larger then there may again be a spiral sink or convergence to a mixture of two strategies (Fig.~\ref{fig5}i).
These diverse dynamical outcomes for the replicator equations with noise agree with Monte-Carlo simulations (see Supplementary Fig.~3).

In summary, our results on three-strategy games reinforce the general conclusions we found with two-strategy games: environmental fluctuations can both strengthen the stability of interior states, with a mixture of strategies, and weaken the stability of boundary states. This phenomenon ultimately promotes a greater diversity of strategies, compared to a noise-free environment. Our analysis extends to other three-strategy games, where we also find that noisy environments produce a greater number of interior equilibria (see Supplementary Fig.~4); and also to four-strategy games characterized by cyclic dominance among strategies, where we find oscillating dynamics (see Supplementary Fig.~5). 


\subsection*{General fitness functions and update rules}
\label{section3.3}

In this section we extend our analysis of behavioral evolution in noisy environments beyond the classic birth-death process and the exponential fitness function. We have focused on the birth-death process when deriving the associated replicator equation. But there are several alternative updating rules that are widely used in evolutionary game theory, including the death-birth process, imitation process, and pairwise comparison. In Section 3.1 of Supplementary Information, we show that noisy environments do not change the dynamics of strategy evolution under the pairwise comparison rule. But for death-birth and imitation processes the effects of environmental noise are the same as for the birth-death process.

We have assumed a standard exponential relationship between payoff and fitness, $f_i=\exp(s\Pi_i)$.
We can extend our analysis to an arbitrary fitness function $f_i=f(s\Pi_i)$, which simultaneously provides some intuition for our results. Under the birth-death process, the replicator equation for an arbitrary fitness function has the form (see Supplementary Information, Section 3.1):
\begin{equation}
    \dot{x}_i=sx_i\left[\delta_1\bar{\sigma}(\bar{\sigma}-\sigma_i)+\frac{\delta_2}{2}(\sigma_i^2-M_2)+(\pi_i-\bar{\pi})\right],
    \label{eq:general system equation}
\end{equation}
where $\bar{\sigma}$ represents the mean intensity of payoff fluctuations in the population (i.e., $\bar{\sigma}=\sum_{i=1}^m x_i \sigma_i$) and $M_2$ represents the second moment (i.e., $M_2=\sum_{i=1}^m x_i \sigma_i^2$) of fluctuation intensities across the population. Here $\delta_1=f'(0)/f(0)$ describes how fast fitness increases with selection intensity and payoff, which is always positive.
And $\delta_2=f''(0)/f'(0)$ measures the convexity  of the fitness function, which is a measure of  risk preference. (In the terminology of economics $-\delta_2$ equals the Arrow-Pratt absolute risk aversion \cite{arrow1971Essays,Pratt1978,Nicholson2012}). When $\delta_1=\delta_2=1$, Eq.~\ref{eq:general system equation} simplifies to the case of an exponential fitness function (Eq.~\ref{eq:multi-strategy system equation}), which has the nice property that strategy dynamics depends only on differences in the intensity of fluctuation between strategies, such as $\sigma_i-\sigma_j$.

For a general fitness function, however, the dynamics of strategy frequencies will depend on the specific values of fluctuation intensities, $\{\sigma_i\}$. In fact, even if the fitness function is linear, e.g. $f_i=1+s \pi_i$, then the $\delta_1$ term above remains non-zero (although $\delta_2=0$), and so there will still be an effect of noise on the evolutionary dynamics.  More generally, Eq.~\ref{eq:general system equation} shows that the effects of environmental fluctuations arise from two distinct factors. One factor is governed by the birth-death updating rule (the first term of Eq.~\ref{eq:general system equation}) and the second factor is determined by the non-linear fitness function (the second term of Eq.~\ref{eq:general system equation}). 

Here we assume that $\sigma_i$ is positive for all strategies $i$. 
The first term of Eq.~\ref{eq:general system equation} implies that a strategy $i$ whose fluctuation intensity ($\sigma_i$) is lower than the average intensity gains an evolutionary advantage. This effect (which arises from the death-birth process itself, see Section~3.1 in Supplementary Information) always provides an advantage to strategies that experience less environmental noise, compared with the average.

The second term of Eq.~\ref{eq:general system equation} arises from non-linearity of the fitness function itself. If the fitness function is concave (i.e., $\delta_2<0$), fluctuations in payoff will produce smaller expected fitness than the deterministic payoff, providing an evolutionary advantage for strategies that experience less noise. On the contrary, if the fitness function is convex (i.e., $\delta_2>0$), then noisy payoffs are advantageous to a behavioral type. And so this second effect can either help or hinder a strategy subject to less environmental noise, depending upon the convexity of the fitness function -- analogous to the effects of risk-seeking versus risk-averse preferences in economics \cite{Pratt1978,arrow1971Essays,wang2024}.


\section*{Discussion}





There is already a rich and varied literature in evolutionary game theory that seeks to explain the diversity of social behaviors observed within and between populations. The core theory, which had its inception with the work of Maynard-Smith \cite{Taylor1978,smith1982evolution}, has since been expanded to account for a huge range of realistic complications, including arbitrary spatial structure. Nonetheless, the theory typically assumes that social interactions bring deterministic payoffs that do not change over time. It is precisely this assumption -- which is often violated in reality -- that we have relaxed in this study.



Our analysis reveals that noisy social environments fundamentally change behavioral outcomes in a population of individuals who imitate successful strategies.  For instance, in the prisoner's dilemma, which is classically associated with only the dominance of defection, an intermediate amount of noise produces a new equilibrium that permits the coexistence of cooperation and defection. This can occur even when perturbations are symmetric, with an equal chance of increasing or decreasing payoffs. Stronger noise can entirely eliminate the equilibrium with pure defection, so that a stable co-existence becomes the only long-term outcome. Systematic study of more complex social interactions, such as three- and four-strategy games, confirms the diversity of  dynamical patterns that noise produces -- typically by increasing the stability of interior equilibria while weakening boundary equilibria. Noisy payoffs can even produce stable, cyclical dynamics that cannot occur in the absence of noise.

Given our results, as well as the ubiquity of real-world
perturbations on the reward structure of social interactions \cite{luhmann2011intolerance,ketchpel1994forming}, it seems likely that diversity of social behavior observed across populations \cite{bouwmeester2017registered, henrich2001search,cox1991effects,gelfand2007cross,tung2008cross,stahl2010unraveling} 
may be attributed, in part, to different levels of noise in different populations. It remains an open and largely empirical question to measure the frequency and structure of stochastic perturbations to payoffs in different populations, but their effects can now be interpreted within the framework of evolutionary game theory.

At the broadest level, our results on noisy social  environments are analogous to a phenomenon called the ``storage effect'' in evolutionary ecology \cite{Chesson1981,Pake1995,Caceres1997,Kelly2002}, 
where environmental fluctuations tend to promote coexistence. In both settings, the essential intuition is that a fluctuating environment does not allow any single species or type to consistently outperform others under all conditions. Each type effectively stores the benefits accrued at earlier, favorable times to withstand periods that are unfavorable. Our work extends this concept to the context of behavioral types, whose fitness also depends on the frequency of alternative types in the population, arising from pairwise social interactions. At the same time, our results are not simply the effect of storage, because we have seen that the consequences of noise also depend on the form of evolutionary updates (e.g., for updates by pairwise comparison, noisy environments have no effect on dynamics, Supplementary Fig.~6).


There are several notable works that address situations related to this study, including work on multiple games \cite{Donahue2020,Su2022,Venkateswaran2019}, stochastic games \cite{Hilbe2018a,Su2019a,Wang2021}, and games with environmental feedback \cite{Weitz2016,Tilman2020}. Those studies differ from this one in motivation and conclusions. Research on multiple games addresses scenarios where individuals can engage in multiple different games simultaneously, leading to different outcomes for the same strategy profiles in different interactions \cite{Hilbe2018a,Su2019a,Wang2021}; nonetheless the outcome is still deterministic for each interaction.
Studies on stochastic games or games with environmental feedback, by contrast, consider potential changes in the environmental state, but the environmental pattern typically  depends on the composition of the population \cite{Hilbe2018a,Su2019a,Wang2021,Weitz2016,Tilman2020}, which is again distinct from the unbiased exogenous perturbations we have studied. Other studies have considered demographic stochasticity, fluctuating rates of reproduction, or observational uncertainty \cite{foster1990stochastic,fudenberg1992evolutionary,traulsen2004stochastic,Constable2016,wang2023a,wang2024}; these forms of randomness apply to each individual independently or to each type independent of the frequencies of other types. By contrast, we have considered a form of ``global'' perturbation to the game payoff structure, which captures the effect of a stochastic change in the social environment for the entire population simultaneously.

There is also prior work on various forms of global noise \cite{houchmandzadeh2012selection,huang2015stochastic,Wienand2017,Ashcroft2014,Meyer2018,Stollmeier2018,asker_coexistence_2023}, which are also known to reverse the direction of evolution in populations \cite{houchmandzadeh2012selection,Meyer2018,Stollmeier2018,Constable2016,wang2023a}.  
In many of these prior studies 
the nature of environmental noise is different than ours, focusing on fluctuations in the carrying capacity \cite{houchmandzadeh2012selection,huang2015stochastic,Constable2016,Wienand2017,Taitelbaum2020}, variation in frequency-independent selection intensity\cite{Constable2016,Assaf2013,asker_coexistence_2023}, 
or periodic changes to payoffs \cite{Stollmeier2018} -- as opposed to random fluctuations in the payoff matrix. That is to say, the noise we study arises from randomness in the nature of strategic interactions themselves, rather than stochasticity in population size, selection intensity, or demographic processes. 

Prior work on global noise has often focused on the fixation probability of a type
\cite{Wienand2017,Assaf2013,Ashcroft2014,Meyer2018}. Our approach, by contrast, also describes dynamical patterns in the interior of state space, which reveals qualitatively new forms of stable co-existence and limit cycles that cannot occur without noise.  
Some work has examined environmental noise by adding a stochastic term to the discrete replicator equation (DRE) but without an individual-based model \cite{Stollmeier2018,zhengEnvironmental2018}. The DRE produces radically different dynamics, including chaos, compared to the classical replicator equation. 

Our study is certainly not without limitations or assumptions made for the sake of simplicity. In the main text, we considered scenarios where fluctuations arise from a single source of white noise. In other words, payoff perturbations arose from a single random variable, and its effects were drawn independently across  different generations. We can, at least, relax these two assumptions.
In Section~3.2 of Supplementary Information, we analyze two additional scenarios. One scenario allows for independent sources of noise affecting each element of the payoff matrix, which still produce novel interior equilibria and foster the coexistence of multiple behaviors (see Supplementary Fig.~9 and Supplementary Fig.~10). The second scenario considers correlated fluctuations across generations, also known as colored noise. 
In this setting, the payoff perturbation in one generation is correlated with the perturbation in the previous generation, with correlation coefficient $1-\nu$ ($0<\nu\le1$). The case $\nu=1$ reduces to white noise, which we have studied above, whereas $\nu<1$ corresponds to colored noise. The value of $\nu$ represents the speed of environmental fluctuation related to that of strategy evolution. 
In Section~3.2, we prove that for a wide range of colored noise ($1/N\ll \nu<1$) the results are qualitatively the same as for white noise (see Section~3.2 of Supplementary Information and Supplementary Fig.~11). 
When the environment fluctuates much more slowly than strategy evolution ($\nu\ll  1/N$), the resulting distribution of behavior is qualitatively different than under white noise, but it still features phenomena that cannot occur in the absence of noise (see Section~3.2 of Supplementary Information, Supplementary Fig.~12, and Supplementary Fig.~13).

Beyond these scenarios, several other extensions warrant exploration. For example, the notion of an evolutionarily stable strategy (ESS) is an essential concept in evolutionary games. It is well understood how an ESS is related,  but not identical, to a stable equilibrium of the classic replicator equation \cite{cressman2014replicator}. The relationship between ESSs and equilibria of the modified replicator equation (Eq.~\ref{eq:replicator}) in a noisy environment will be more complex and remains an open question.
Finally, investigating systems that incorporate both noisy payoffs as well as environmental feedback \cite{Tilman2020}, where the population composition may even influence the intensity of environmental fluctuations, may yield yet greater diversity of dynamical outcomes, or modify the outcomes we have observed in the absence of feedback. These remain outstanding areas for future research.

\section*{Methods}
\subsection*{A replicator equation with environmental noise}
Here we briefly outline the derivation of the modified replicator equation for noisy environments, with details deferred in Supplementary Information. For a game with $m$ strategies, the payoff structure is given by a matrix $\mathbf{A}$ with elements shown in Eq.~\ref{eq: general payoff elements}. We denote the frequency of individuals using strategy $i$ by $x_i$, and the payoff for such individuals is given by
\begin{equation}
    \Pi_i=\sum_{j\le m}A_{ij}x_j=\sum_{j\le m}a_{ij}x_j+\frac{\xi}{\sqrt{s}}\sum_{j\le m}\sigma_{ij}x_j=\pi_i+\frac{\xi}{\sqrt{s}}\sigma_i.
\end{equation}
An individual's payoff determines his fitness, or reproductive capacity, according to the function$f_i=\exp(s\Pi_i)$ \cite{Wu2013,Hauert2012106,Gokhale20105500}. Given the environment state $\xi=\delta$, the probability that $x_i$ will increase or decreases by $1/N$ in each generation (for the birth-death process) is
\begin{subequations}
    \begin{align}
        &T^+_i(\mathbf{x}|\xi=\delta)=\frac{x_if_i^\delta}{\sum_{j=1}^m x_jf_j^\delta}(1-x_i), \\
        &T^-_i(\mathbf{x}|\xi=\delta)=\frac{\sum_{x\ne i}x_jf_j^\delta}{\sum_{j=1}^m x_jf_j^\delta}x_i,
    \end{align}
\end{subequations}
where $f_i^\delta=\exp(\pi_i+\delta \sigma_i/\sqrt{s})$ and $\mathbf{x}=[x_1,x_2,\cdots,x_m]^T$. Applying the law of total probability, the probability that $x_i$ increases or decreases by $1/N$ in each generation is
\begin{subequations}
    \begin{align}
        &T^+_i(\mathbf{x})=\mathbb{E}\left[T^+(x|\xi)\right]=\int T^+(x|\xi=\delta) p(\delta){\rm d}\delta, \\
        &T^-_i(\mathbf{x})=\mathbb{E}\left[T^-(x|\xi)\right]=\int T^-(x|\xi=\delta) p(\delta){\rm d}\delta,
    \end{align}
\end{subequations}
where $p(\delta)$ is the probability density function of $\xi$. The expected change of $x$ in each generation is therefore $\frac{1}{N}[T^+_i(\mathbf{x})-T^-_i(\mathbf{x})]$. We rescale the time by setting $t=\tau/N$ where $\tau$ represents generations. Then, the time derivative of $x_i$ is (see detailed description in Supplementary Information)
\begin{equation}
    \dot{x}_i=\frac{\frac{1}{N}[T^+_i(\mathbf{x})-T^-_i(\mathbf{x})]}{\frac{1}{N}}=T^+_i(\mathbf{x})-T^-_i(\mathbf{x}).
\end{equation}
Expanding $T^+_i$ and $T^-_i$ in a Taylor series and truncating at first order in $s$ produces the ODE system stated in Eq.~\ref{eq:multi-strategy system equation}.

\subsection*{Emergence of limit cycles in three-strategy games}
For rock-paper-scissor games, the frequencies of the three strategies are denoted by $x$, $y$, and $z$ respectively. Given the relation $x+y+z=1$, we choose $x$ and $y$ to be the two free variables. The  dynamics of $x$ and $y$ are then given by Eq.~\ref{eq:multi-strategy system equation}. In the case of $\sigma_{ij}=ka_{ij}$, the system contains only one interior equilibrium point, $e^*=(x^*,y^*)=(\frac{1}{3},\frac{1}{3})$. The two eigenvalues of the Jacobian at this equilibrium are given by
\begin{equation}
    \lambda_{1,2}=\frac{s(\alpha-\beta)}{6}\pm 
    i\frac{s\sqrt{3}}{6}(\alpha+\beta).
\end{equation}
Here $i$ is the imaginary unit. Denote $\alpha-\beta=\mu$. If $\mu>0$, $e^*$ is always unstable and the system exhibits a spiral source in the neighborhood of $e^*$. If $\mu<0$, $e^*$ is stable and thus a spiral sink. As $\mu$ passes through $0$ the system exhibits an Andronov-Hopf bifurcation. To check whether it can produce a limit cycle we compute the first Lyapunov exponent $l_1(0)$.  According to \cite{Guckenheimer1983}, we have
\begin{equation}
    l_1(0)=-\frac{s}{2}\alpha^2k^2.
\end{equation}
Thus, if $k>0$, meaning there is some noise, the first Lyapunov exponent is negative which admits a supercritical Andronov-Hopf bifurcation and a stable limit cycle emerges for $\mu>0$ (negative-sum RPS). When $k=0$, meaning a deterministic environment, the bifurcation is degenerate, and so no limit cycles exist.




\bibliographystyle{sn_nature}
\bibliography{reference}

\includepdf[pages=-]{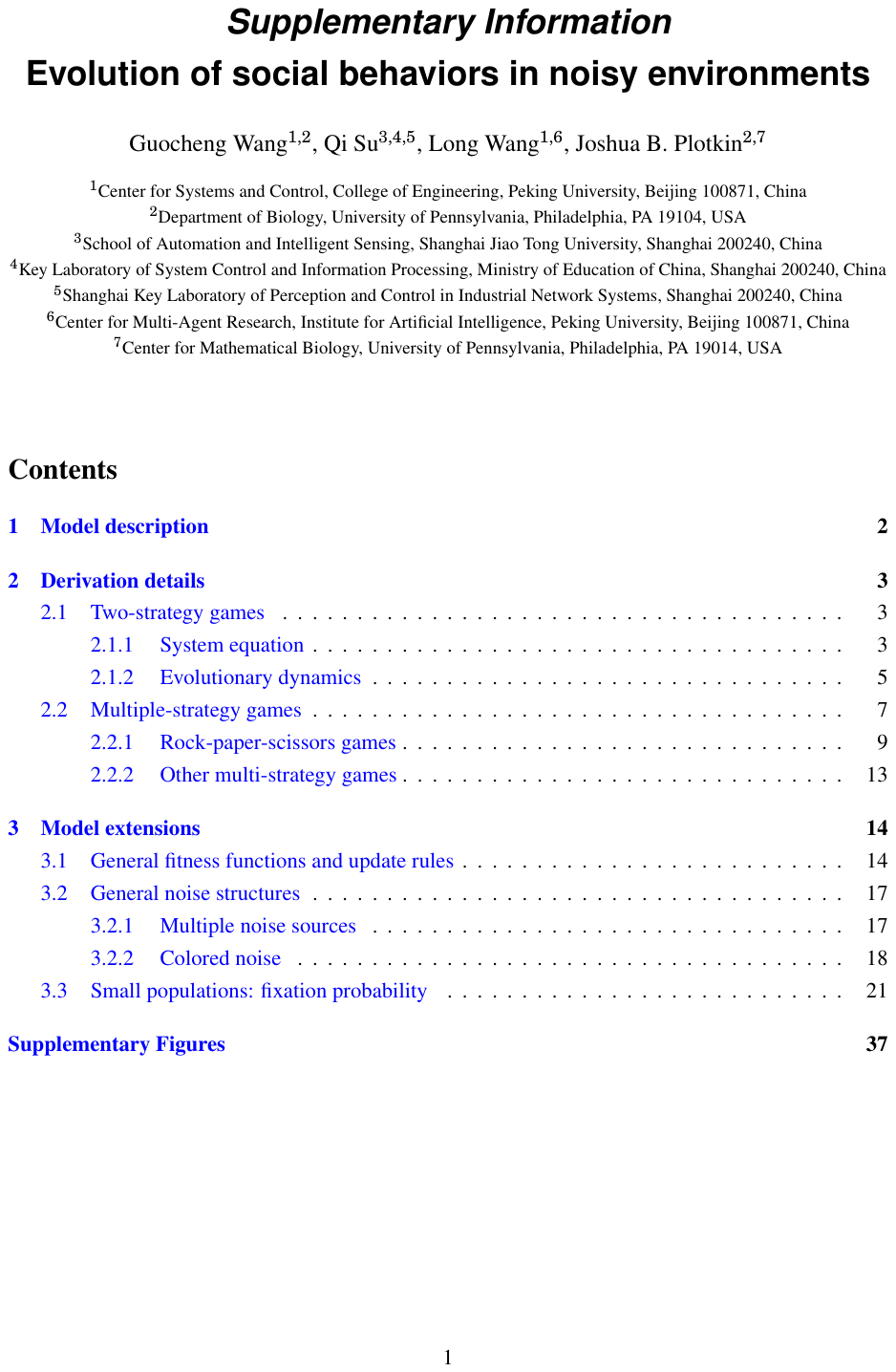}

\end{document}